\documentclass[]{aa}
\usepackage{times}
\usepackage{graphics}

\begin{document}

   \thesaurus{09          
              (02.03.3;   
               02.08.1;   
               02.12.1;   
               06.01.1;   
               06.07.2;   
               06.16.2)}   

   \title{Line formation in solar granulation}
   \subtitle{III. The photospheric Si and meteoritic Fe abundances}

   \author{M. Asplund\inst{1,2}
          }


   \institute{
              NORDITA, 
              Blegdamsvej 17, 
              DK-2100 ~Copenhagen {\O}, 
              Denmark      
              \and
              present address: Uppsala Astronomical Observatory,
              Box 515,
              SE-751 20 ~Uppsala,
              Sweden (e-mail: martin@astro.uu.se)
              }

   \date{Received: January 24, 2000; accepted: May 4, 2000 }

\authorrunning{M. Asplund}
\titlerunning{Solar line formation: III. The photospheric Si abundance}

   \maketitle

   \begin{abstract}
Using realistic hydrodynamical simulations of the solar surface
convection as 3D, time-dependent, 
inhomogeneous model atmospheres, the solar photospheric Si
abundance has been determined to be
log\,$\epsilon_{\rm Si} = 7.51\pm0.04$.
This constitutes a difference
of 0.04\,dex compared with previous estimates based on the
1D Holweger-M\"uller (1974) model, of which half is attributable to the
adopted model atmosphere and the remaining part
to the improved quantum mechanical broadening treatment.
As a consequence, all meteoritic abundances should be 
adjusted downwards by the same amount. 
In particular the meteoritic Fe abundance will be
log\,$\epsilon_{\rm Fe} = 7.46\pm0.01$, 
in good agreement with the recently determined photospheric Fe
abundance (Asplund et al. 2000b). 
The existing uncertainties unfortunately prevent an observational
confirmation of the postulated effects of elemental migration of metals 
in the Sun.

      \keywords{Convection -- Hydrodynamics --  
                Line: formation -- Sun: abundances --
                Sun: granulation -- Sun: photosphere 
               }
   \end{abstract}

\section{Introduction \label{s:intro}}

To compare the meteoritic and solar photospheric chemical compositions
it is necessary to have a common reference element. 
The honour is normally given to Si, which is an abundant element
and a natural choice to measure other elemental abundances against
in meteorites considering the volatility of hydrogen. In order to place 
all other meteoritic abundances on an absolute scale to confront
with the photospheric abundances, one must therefore 
accurately know the photospheric Si/H ratio. As a consequence all
meteoritic absolute abundances 
will depend on the measured solar photospheric Si abundance. 
In practice, one often utilizes additional elements besides Si
to anchor the two scales more firmly to each other
(e.g. Anders \& Grevesse 1989).

Several determinations of the solar photospheric Si abundance exist in the
literature (e.g. Holweger 1973; Lambert \& Luck 1978; Becker et al. 1980)
using the Holweger-M\"uller semi-empirical model atmosphere
(Holweger \& M\"uller 1974) and the use of equivalent widths and
the microturbulence concept. Given the recent progress in 
the construction of self-consistent, 3D, hydrodynamical simulations
of the solar surface convection
(e.g. Stein \& Nordlund 1989, 1998; Asplund et al. 2000a,b, 
hereafter Paper I and II)
and the significance of the solar photospheric Si abundance, a re-analysis
seemed warranted, which is presented here. These convection simulations have  
now reached a very high degree of realism. Without relying on
any free parameters, such as the mixing length parameters, the 
simulations successfully reproduce the granulation topology and
statistics (Stein \& Nordlund 1998), 
constraints from helioseismology (Rosenthal et al. 1999) 
and detailed spectral line shapes, shifts and asymmetries
(Paper I and II).  
In particular, line profiles are accurately described without invoking
any micro- and macroturbulence which are necessary in classical
1D spectral analyses. As a consequence several possible sources
of uncertainties (e.g. model atmosphere limitations, equivalent widths and
microturbulence) can be removed when deriving elemental abundances,
which should result in more secure determinations. 

\section{3D spectral line formation and atomic data}

The procedure is identical to that adopted for a recent determination
of the solar photospheric Fe abundance (Paper II).
In particular
the model atmosphere is a realistic 3D hydrodynamical simulation
of the solar surface convection consisting of 50\,min solar time. 
For further details of the 
simulation and the 3D spectral line formation 
the interested reader is referred to  Paper I and II.

\begin{table}[t!]
\caption{The adopted line data for the Si\,{\sc i} and
Si\,{\sc ii} lines
\label{t:lines}
}
\begin{tabular}{lccccc} 
 \hline
Species & Wavelength & $\chi_{\rm l}$ & log\,$gf$ & $W_\lambda^{\rm a}$ &
log\,$\epsilon_{\rm Si}$ \\
        &    [nm]    & [eV]           &       & [pm]     & \\
\hline \\
Si\,{\sc i} &        564.56130 &  4.93 & -2.04 &  3.40 &  7.52 \\
Si\,{\sc i} &        566.55550 &  4.92 & -1.94 &  4.00 &  7.45 \\
Si\,{\sc i} &        568.44840 &  4.95 & -1.55 &  6.00 &  7.47 \\
Si\,{\sc i} &        569.04250 &  4.93 & -1.77 &  5.20 &  7.51 \\
Si\,{\sc i} &        570.11040 &  4.93 & -1.95 &  3.80 &  7.48 \\
Si\,{\sc i} &        570.84000 &  4.95 & -1.37 &  7.80 &  7.48 \\
Si\,{\sc i} &        577.21460 &  5.08 & -1.65 &  5.40 &  7.51 \\
Si\,{\sc i} &        578.03840 &  4.92 & -2.25 &  2.60 &  7.52$^{\rm b}$ \\
Si\,{\sc i} &        579.30730 &  4.93 & -1.96 &  4.40 &  7.55 \\
Si\,{\sc i} &        579.78560 &  4.95 & -1.95 &  4.00 &  7.55 \\
Si\,{\sc i} &        594.85410 &  5.08 & -1.13 &  8.60 &  7.48 \\
Si\,{\sc i} &        674.16280 &  5.98 & -1.65 &  1.60 &  7.55 \\
Si\,{\sc i} &        697.65130 &  5.95 & -1.07 &  4.30 &  7.56$^{\rm b}$ \\
Si\,{\sc i} &        703.49010 &  5.87 & -0.78 &  6.70 &  7.54$^{\rm b}$ \\
Si\,{\sc i} &        722.62080 &  5.61 & -1.41 &  3.60 &  7.51 \\
Si\,{\sc i} &        768.02660 &  5.86 & -0.59 &  9.80 &  7.54 \\
Si\,{\sc i} &        791.83840 &  5.95 & -0.51 &  9.50 &  7.50$^{\rm b}$ \\
Si\,{\sc i} &        793.23480 &  5.96 & -0.37 &  9.70 &  7.53$^{\rm b}$ \\
Si\,{\sc i} &        797.03070 &  5.96 & -1.37 &  3.20 &  7.59 \\
Si\,{\sc ii} &       634.71090 &  8.12 &  0.30 &  5.60 &  7.48 \\
Si\,{\sc ii} &       637.13710 &  8.12 & -0.00 &  3.60 &  7.43 \\
\hline 
\end{tabular}
\begin{list}{}{}
\item[$^{\rm a}$] From Holweger (1973). Note that $W_\lambda$ is only 
listed here to allow easy identification in Fig. \ref{f:si_w} and is not
used for the abundance determinations
\item[$^{\rm b}$] Lines which are entered with half weight into the
final abundance estimate due to uncertainties in broadening, oscillator
strengths or suspected blends
\end{list}

\end{table}

\begin{figure}[t]
\resizebox{\hsize}{!}{\includegraphics{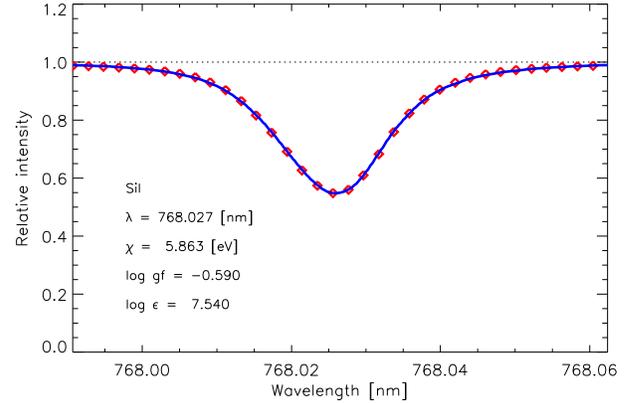}}
\caption{The spatially and temporally averaged 
(diamonds) Si\,{\sc i} 768.0\,nm line calculated
using the 3D inhomogeneous model atmosphere compared with the
observed profile (solid). 
The predicted profile has been convolved with a sinc-function to account
for the finite spectral resolution of the solar atlas 
(Paper I and II).
Since no free parameters but the Si abundance enter
the spectral synthesis, the good agreement supports the conclusion 
that the macroturbulence concept required in classical 1D analyses 
is not necessary when properly accounting for the convective Doppler shifts
in the hydrodynamical simulations}
         \label{f:sii}
\end{figure}

The Si lines used for the study are the same as in Holweger (1973)
and are presented in Table \ref{t:lines} together with the necessary
atomic data. The sample consists of
19 Si\,{\sc i} and 2 Si\,{\sc ii} lines. The necessary transition
probabilities are taken from Becker et al. (1980), which are
the same as measured by Garz (1973) but corrected using more
recent lifetime measurements. 
We note that 
the lifetime measurements of Smith et al. (1987) in fact support
the scale of Garz (1973) but until the conflicting lifetime situation
has been clarified, I have chosen to adopt the values of Becker et al. (1980)
to emphasize the effect of using 3D model atmospheres;
new improved measurements of the branching ratios and lifetimes of the
Si lines seem justified.
The collisional broadening data for both Si\,{\sc i} and Si\,{\sc ii} lines 
have been computed following the recipes of Anstee \& O'Mara (1991, 1995),
Barklem \& O'Mara (1997), Barklem et al. (1998) and Barklem (2000, private
communication).
If one instead would embrace the  
classical collisional treatment of Uns\"old (1955) without any
damping enhancement factors the final Si abundance would be 0.02\,dex
higher with a slightly increased scatter. 
Stark broadening and radiative broadening 
was included with data taken from the VALD database (Kupka et al. 1999),
although Stark broadening is only of some significance for 
Si\,{\sc i} 730.49 and 768.03\,nm. 
The Si abundance has been determined from profile fitting of the
individual Si lines through a $\chi^2$-analysis in a similar fashion 
to the study of Nissen et al. (2000)
rather than equivalent widths.   
The solar atlas of Brault \& Neckel (1987)
and Neckel (1999) provided the observed spectrum. 
Fig. \ref{f:sii} shows an example of the achieved agreement
between predicted and observed line profiles. It should be emphasized
that this has been accomplished without invoking any free
parameters besides the elemental abundance; the result is clearly
superior to what is possible using 1D model atmospheres such as the
Holweger-M\"uller (1974) model.

\section{The solar photospheric Si abundance
\label{s:si}}

\begin{figure}[t]
\resizebox{\hsize}{!}{\includegraphics{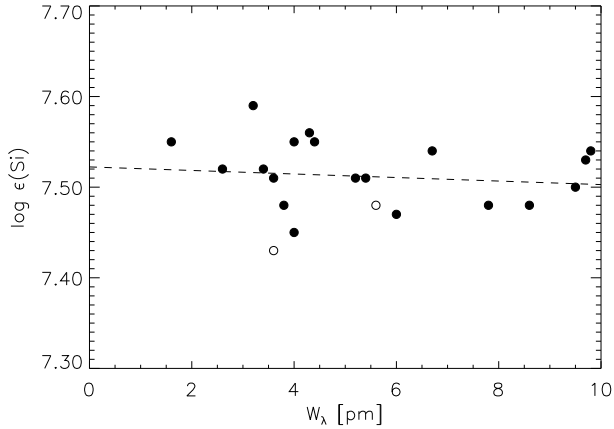}}
\caption{The individual Si abundances derived from the profiles
of Si\,{\sc i} (solid circles)
and Si\,{\sc ii} (open circles) lines as a function of the equivalent
widths (taken from Holweger 1973); it should be emphasized however that
equivalent widths have not been used for the abundance determinations. 
The dashed line represents a
least square fit to the Si abundances. The lack of an apparent trend
supports the conclusion that no microturbulent velocity is
necessary to include in the spectral synthesis when correctly
including the Doppler shifts inherent in the convection simulations}
         \label{f:si_w}
\end{figure}

Fig. \ref{f:si_w} reveals no clear trend in the derived individual
Si abundances with line strength; the span in excitation potential
is unfortunately too small to clearly delineate any possible
trend. When including all Si\,{\sc i}
and Si\,{\sc ii} lines with equal weight,
the estimated solar photospheric Si abundance is
log$\,\epsilon_{\rm Si} = 7.51\pm0.04$
\footnote{On the customary logarithmic abundance scale defined
to have log$\,\epsilon_{\rm H}=12.00$}, where the quoted uncertainty
is the standard deviation. 
If equivalent widths from the literature
(Holweger 1973) had been used instead of profile fitting, the
mean abundance would have remained the same but the scatter would have
increased somewhat. 
When comparing line profiles 
it was noticed that a few lines are less suitable
for abundance determinations.
The most likely reasons for the poor agreement in the profile
fitting appear to be suspected blends or erroneous broadening 
treatment. Such lines enter the final abundance estimate with
half weight as marked in Table \ref{t:lines}. 

The re-analysis of the solar Si lines then results in 
$${\rm log}\,\epsilon_{\rm Si} = 7.51\pm0.04,$$
which is only slightly less than the value ($7.55\pm0.05$)
published by Becker et al. (1980)
with the same set of transition probabilities. 
Half of this difference is due to
the adoption of a 3D hydrodynamical model atmosphere instead of
the Holweger-M\"uller (1974) model atmosphere and the remaining part
to the improved quantum mechanical broadening treatment instead of
Uns\"old (1955) pressure broadening.  
Clearly the Si lines are little sensitive to the detailed atmospheric
structure and emphasizes that the main uncertainty is likely to be
the absolute $gf$-scale, although the small scatter suggests that the
relative measurements by Garz (1973) are of good internal consistency.
Departures from LTE will probably not significantly influence the
derived mean abundance, considering the relatively small scatter for
the Si\,{\sc i} lines and since ionization balance seems to
be essentially fulfilled; some caution in this matter should still 
be exercised however, since only two Si\,{\sc ii} lines enter the analysis.
Additional Si\,{\sc ii}
lines would certainly be very helpful in securing the results further.

\section{The photospheric and meteoritic abundance scales}

A re-analysis of the solar photospheric Si abundance using 3D, time-dependent,
inhomogeneous model atmospheres have revealed only a difference of
0.04\,dex (10\%) with
previously published estimates. As described in Sect. \ref{s:intro}
this also suggests that the meteoritic abundances remain largely
unaltered compared with the commonly adopted values
(Grevesse \& Sauval 1998). 
If only Si is used to place the meteoritic and photospheric abundances
on a common scale, the meteoritic Fe abundance for example will be
log$\,\epsilon_{\rm Fe} = 7.46\pm0.01$ 
(an additional 0.01\,dex decrease may come from the existing difference in 
adopted meteoritic and photospheric Si abundances, Grevesse \& Sauval 1998). 
It should be noted though that 
several other elements are often utilized as well
(Na, Mg, Si, Ca, V, Cr, Co, Ni, Y, Zr, Nb and Mo, Anders \& Grevesse 1989) 
for the purpose
and it is not clear whether all these elements will behave similarly
to Si when changing from the Holweger-M\"uller (1974)
model atmosphere to the 3D solar convection simulations forming the
basis of this work. However, given the similarity in this respect
between Si and Fe 
(Paper II) it seems reasonable that all
meteoritic abundances indeed need to be adjusted downwards
by about 0.04\,dex.

The difference between the meteoritic Fe abundance arrived at here 
and the recently determined photospheric abundance 
(Paper II) is only 0.02\,dex 
for Fe\,{\sc i} (log$\,\epsilon_{\rm Fe} = 7.44\pm0.05$)
and 0.01\,dex for Fe\,{\sc ii} (log$\,\epsilon_{\rm Fe} = 7.45\pm0.10$)
lines, when solely relying on the Si results presented here
for the absolute scale for the meteoritic abundances.
It is worthwhile to keep in mind, however, 
that the photospheric Fe results using
3D model atmospheres are based on the assumption of LTE. For some
1D model atmospheres there are indications that low-excitation 
Fe\,{\sc i} lines may be influenced by departures from LTE at the
level of $\Delta {\rm log} \epsilon_{\rm Fe} \la 0.1$\,dex
(Solanki \& Steenbock 1988; Shchukina 2000, private communication),
which would, if assumed to be valid also in the 3D case, degrade the
agreement between the 3D LTE results for Fe\,{\sc i} and Fe\,{\sc ii}
(Paper II). However, until detailed 3D NLTE calculations have been
performed such a conclusion appears to be 
premature since the NLTE results are sensitive
to the adopted model atmosphere. Rather the consistency between the
meteoritic value and the 
abundances derived from both weak and strong Fe\,{\sc i} and 
Fe\,{\sc ii} lines and the absence of any trend in abundances with 
excitation potential may be interpreted as NLTE effects being
insignificant; in any case the Fe\,{\sc ii} results should be robust against
departures from LTE 
(Solanki \& Steenbock 1988; Shchukina 2000, private communication). 

The difference between the here estimated meteoritic and
photospheric Fe abundances
is slightly smaller than the value of 0.04\,dex (10\%) 
expected due to diffusion and gravitational settling 
(e.g. Vauclair 1998), similarly
to the situation for He which has been
uncovered by helioseismology (e.g. Basu 1998).
However, the absolute transition probabilities for
Fe and Si are unfortunately still not of sufficient accuracy 
to definitely reveal such a small effect,
even though the atmospheric modelling now has answered the challenge. 
Furthermore, and more importantly, the absolute abundances in 
meteorites have been obtained by enforcing the average differences
between the photospheric and meteoritic abundances for some 10 metals
to vanish. It is therefore not surprising that the average difference
in abundances for the 59 elements (excluding e.g.
Li, Be, B, which may have been depleted) is $<0.01$\,dex 
(Grevesse \& Sauval 1998). 
Thus, in order to observationally prove elemental migration 
for the Sun through spectroscopy, 
{\it differential} diffusion effects must be identified, which are
expected to be significantly less than the {\it overall} 
depletion of metals by about 10\%.

\begin{acknowledgements}
Nicolas Grevesse is thanked for helpful discussions regarding
solar and meteoritic abundances.
The author is greatly 
indebted to Paul Barklem for computing quantum mechanical broadening data
specifically for the Si lines.
Discussions with Natalia Shchukina regarding departures from
LTE for Fe are much appreciated, as are  
the constructive suggestions by an anonymous referee..  
The VALD database has been providing an abundance of useful data, 
which is gratefully acknowledged by the author. 
\end{acknowledgements}


\end{document}